 \documentclass[twocolumn]{aastex62}
\pdfoutput=1 
\usepackage{amsmath,amstext}
\usepackage[T1]{fontenc}
\usepackage{apjfonts} 
\usepackage[figure,figure*]{hypcap}
\usepackage[utf8]{inputenc}
\usepackage{array}
\usepackage{makecell}
\usepackage{booktabs}
\usepackage{threeparttable}
\usepackage{mathrsfs}
\usepackage{mathtools}
\usepackage{footmisc}
\usepackage{multirow}

\newcommand*\diff{\mathop{}\!\mathrm{d}}

\shorttitle{IceCube-ANITA point source search}
\shortauthors{M. Aartsen et al.}

\begin{document}
\email{analysis@icecube.wisc.edu}
\title{A search for IceCube events in the direction of ANITA neutrino candidates}

\affiliation{III. Physikalisches Institut, RWTH Aachen University, D-52056 Aachen, Germany}
\affiliation{Department of Physics, University of Adelaide, Adelaide, 5005, Australia}
\affiliation{Dept. of Physics and Astronomy, University of Alaska Anchorage, 3211 Providence Dr., Anchorage, AK 99508, USA}
\affiliation{Dept. of Physics, University of Texas at Arlington, 502 Yates St., Science Hall Rm 108, Box 19059, Arlington, TX 76019, USA}
\affiliation{CTSPS, Clark-Atlanta University, Atlanta, GA 30314, USA}
\affiliation{School of Physics and Center for Relativistic Astrophysics, Georgia Institute of Technology, Atlanta, GA 30332, USA}
\affiliation{Dept. of Physics, Southern University, Baton Rouge, LA 70813, USA}
\affiliation{Dept. of Physics, University of California, Berkeley, CA 94720, USA}
\affiliation{Lawrence Berkeley National Laboratory, Berkeley, CA 94720, USA}
\affiliation{Institut f{\"u}r Physik, Humboldt-Universit{\"a}t zu Berlin, D-12489 Berlin, Germany}
\affiliation{Fakult{\"a}t f{\"u}r Physik {\&} Astronomie, Ruhr-Universit{\"a}t Bochum, D-44780 Bochum, Germany}
\affiliation{Universit{\'e} Libre de Bruxelles, Science Faculty CP230, B-1050 Brussels, Belgium}
\affiliation{Vrije Universiteit Brussel (VUB), Dienst ELEM, B-1050 Brussels, Belgium}
\affiliation{Dept. of Physics, Massachusetts Institute of Technology, Cambridge, MA 02139, USA}
\affiliation{Dept. of Physics and Institute for Global Prominent Research, Chiba University, Chiba 263-8522, Japan}
\affiliation{Dept. of Physics and Astronomy, University of Canterbury, Private Bag 4800, Christchurch, New Zealand}
\affiliation{Dept. of Physics, University of Maryland, College Park, MD 20742, USA}
\affiliation{Dept. of Astronomy, Ohio State University, Columbus, OH 43210, USA}
\affiliation{Dept. of Physics and Center for Cosmology and Astro-Particle Physics, Ohio State University, Columbus, OH 43210, USA}
\affiliation{Niels Bohr Institute, University of Copenhagen, DK-2100 Copenhagen, Denmark}
\affiliation{Dept. of Physics, TU Dortmund University, D-44221 Dortmund, Germany}
\affiliation{Dept. of Physics and Astronomy, Michigan State University, East Lansing, MI 48824, USA}
\affiliation{Dept. of Physics, University of Alberta, Edmonton, Alberta, Canada T6G 2E1}
\affiliation{Erlangen Centre for Astroparticle Physics, Friedrich-Alexander-Universit{\"a}t Erlangen-N{\"u}rnberg, D-91058 Erlangen, Germany}
\affiliation{Physik-department, Technische Universit{\"a}t M{\"u}nchen, D-85748 Garching, Germany}
\affiliation{D{\'e}partement de physique nucl{\'e}aire et corpusculaire, Universit{\'e} de Gen{\`e}ve, CH-1211 Gen{\`e}ve, Switzerland}
\affiliation{Dept. of Physics and Astronomy, University of Gent, B-9000 Gent, Belgium}
\affiliation{Dept. of Physics and Astronomy, University of California, Irvine, CA 92697, USA}
\affiliation{Karlsruhe Institute of Technology, Institut f{\"u}r Kernphysik, D-76021 Karlsruhe, Germany}
\affiliation{Dept. of Physics and Astronomy, University of Kansas, Lawrence, KS 66045, USA}
\affiliation{SNOLAB, 1039 Regional Road 24, Creighton Mine 9, Lively, ON, Canada P3Y 1N2}
\affiliation{Department of Physics and Astronomy, UCLA, Los Angeles, CA 90095, USA}
\affiliation{Department of Physics, Mercer University, Macon, GA 31207-0001, USA}
\affiliation{Dept. of Astronomy, University of Wisconsin, Madison, WI 53706, USA}
\affiliation{Dept. of Physics and Wisconsin IceCube Particle Astrophysics Center, University of Wisconsin, Madison, WI 53706, USA}
\affiliation{Institute of Physics, University of Mainz, Staudinger Weg 7, D-55099 Mainz, Germany}
\affiliation{Department of Physics, Marquette University, Milwaukee, WI, 53201, USA}
\affiliation{Institut f{\"u}r Kernphysik, Westf{\"a}lische Wilhelms-Universit{\"a}t M{\"u}nster, D-48149 M{\"u}nster, Germany}
\affiliation{Bartol Research Institute and Dept. of Physics and Astronomy, University of Delaware, Newark, DE 19716, USA}
\affiliation{Dept. of Physics, Yale University, New Haven, CT 06520, USA}
\affiliation{Dept. of Physics, University of Oxford, Parks Road, Oxford OX1 3PU, UK}
\affiliation{Dept. of Physics, Drexel University, 3141 Chestnut Street, Philadelphia, PA 19104, USA}
\affiliation{Physics Department, South Dakota School of Mines and Technology, Rapid City, SD 57701, USA}
\affiliation{Dept. of Physics, University of Wisconsin, River Falls, WI 54022, USA}
\affiliation{Dept. of Physics and Astronomy, University of Rochester, Rochester, NY 14627, USA}
\affiliation{Oskar Klein Centre and Dept. of Physics, Stockholm University, SE-10691 Stockholm, Sweden}
\affiliation{Dept. of Physics and Astronomy, Stony Brook University, Stony Brook, NY 11794-3800, USA}
\affiliation{Dept. of Physics, Sungkyunkwan University, Suwon 16419, Korea}
\affiliation{Institute of Basic Science, Sungkyunkwan University, Suwon 16419, Korea}
\affiliation{Dept. of Physics and Astronomy, University of Alabama, Tuscaloosa, AL 35487, USA}
\affiliation{Dept. of Astronomy and Astrophysics, Pennsylvania State University, University Park, PA 16802, USA}
\affiliation{Dept. of Physics, Pennsylvania State University, University Park, PA 16802, USA}
\affiliation{Dept. of Physics and Astronomy, Uppsala University, Box 516, S-75120 Uppsala, Sweden}
\affiliation{Dept. of Physics, University of Wuppertal, D-42119 Wuppertal, Germany}
\affiliation{DESY, D-15738 Zeuthen, Germany}

\author{M. G. Aartsen}
\affiliation{Dept. of Physics and Astronomy, University of Canterbury, Private Bag 4800, Christchurch, New Zealand}
\author{M. Ackermann}
\affiliation{DESY, D-15738 Zeuthen, Germany}
\author{J. Adams}
\affiliation{Dept. of Physics and Astronomy, University of Canterbury, Private Bag 4800, Christchurch, New Zealand}
\author{J. A. Aguilar}
\affiliation{Universit{\'e} Libre de Bruxelles, Science Faculty CP230, B-1050 Brussels, Belgium}
\author{M. Ahlers}
\affiliation{Niels Bohr Institute, University of Copenhagen, DK-2100 Copenhagen, Denmark}
\author{M. Ahrens}
\affiliation{Oskar Klein Centre and Dept. of Physics, Stockholm University, SE-10691 Stockholm, Sweden}
\author{C. Alispach}
\affiliation{D{\'e}partement de physique nucl{\'e}aire et corpusculaire, Universit{\'e} de Gen{\`e}ve, CH-1211 Gen{\`e}ve, Switzerland}
\author{K. Andeen}
\affiliation{Department of Physics, Marquette University, Milwaukee, WI, 53201, USA}
\author{T. Anderson}
\affiliation{Dept. of Physics, Pennsylvania State University, University Park, PA 16802, USA}
\author{I. Ansseau}
\affiliation{Universit{\'e} Libre de Bruxelles, Science Faculty CP230, B-1050 Brussels, Belgium}
% ...
\date{\today}

\collaboration{IceCube Collaboration}
\noaffiliation

\begin{abstract}
During the first three flights of the Antarctic Impulsive Transient Antenna (ANITA) experiment, the collaboration detected several neutrino candidates. Two of these candidate events were consistent with an ultra-high-energy upgoing air shower and compatible with a tau neutrino interpretation. A third neutrino candidate event was detected in a search for Askaryan radiation in the Antarctic ice, although it is also consistent with the background expectation. The inferred emergence angle of the first two events is in tension with IceCube and ANITA limits on isotropic cosmogenic neutrino fluxes. Here, we test the hypothesis that these events are astrophysical in origin, possibly caused by a point source in the reconstructed direction. Given that any ultra-high-energy tau neutrino flux traversing Earth should be accompanied by a secondary flux in the TeV--PeV range, we search for these secondary counterparts in seven years of IceCube data using three complementary approaches. In the absence of any significant detection, we set upper limits on the neutrino flux from potential point sources. We compare these limits to ANITA’s sensitivity in the same direction and show that an astrophysical explanation of these anomalous events under Standard Model assumptions is severely constrained \textcolor{black}{regardless of source spectrum.}
\end{abstract}

\section{Introduction}
\label{sec:intro}
Ever since the detection of high-energy neutrinos of cosmic origin by IceCube in 2013 \citep{Aartsen:2013jdh}, experiments and theoreticians alike have continued to probe the nonthermal processes in the Universe to understand their origins. The bulk of these astrophysical neutrinos are believed to be created in hadronic interactions between cosmic rays and ambient matter or radiation fields in the vicinity of cosmic accelerators \citep{Gaisser:1994yf}, and  their detections can be used to point back to the acceleration sites. Although the first evidence of a neutrino point source, the blazar TXS 0506+056, was reported in 2018 \citep{IceCube:2018cha, IceCube:2018dnn}, the overwhelming majority of the measured neutrino flux remains unexplained. 

Additionally, another population of neutrinos could exist at extremely high energies. Cosmogenic neutrinos are believed to be the result of interactions between ultra-high-energy (UHE) cosmic rays with the cosmic microwave background (CMB) \citep{Greisen:1966jv, Zatsepin:1966jv}. This population is expected to manifest as an isotropic flux at Earth, as cosmic ray primaries can travel outside the vicinity of their accelerators before interacting with the CMB. 

The Antarctic Impulsive Transient Antenna (ANITA) project is a balloon experiment, designed with the primary purpose of detecting the UHE cosmogenic neutrino flux \citep{Gorham:2008dv, Hoover:2010qt, Allison:2018cxu}. Although this is the project's primary scientific goal, the experiment is sensitive to a wide array of impulsive radio signals, and ANITA's first three flights have resulted in a few interesting detections. In this work, we focus on three events observed by ANITA in its searches, all of which have potential neutrino interpretations. Throughout this work, we refer to and explore these events as ``neutrino candidates.'' In the third flight, one Askaryan neutrino candidate (AAC) event was simultaneously identified in one analysis searching for Askaryan emission \citep{Askaryan:1962hbi} and was found to be subthreshold in another. This Earth-skimming event has a signal shape consistent with impulsive broadband emission characteristic of a neutrino origin, and it also came from a location on the continent consistent with simulated distribution of neutrinos of all flavors \citep{Allison:2018cxu}. However, the detection of one candidate event is consistent with the background level estimates of $0.7^{+0.5}_{-0.3}$ for these analyses. ANITA also reported two additional events, each consistent with an astrophysical $\nu_{\tau}$ emerging from Earth \citep{Gorham:2016zah, Gorham:2018ydl}. In this scenario, a $\nu_{\tau}$ undergoes a charged-current interaction (CC) with a nucleus in Earth. The $\tau$-lepton produced in this interaction subsequently decays in the atmosphere, producing an extensive air shower (EAS). The polarity of the radio signal makes it possible to identify and reject downward moving cosmic-ray--induced EAS, as the radio signals of these EAS acquire a phase reversal (opposite polarity) from reflection off the Antarctic ice, while an upgoing $\tau$-induced EAS does not acquire this phase reversal. For a complete list of details of these events, see Table \ref{tab:candidates}. 

\begin{table*}
\centering
\begin{threeparttable}
	\caption{Properties of the neutrino candidate events from the first three flights of ANITA, from \citep{Allison:2018cxu, Gorham:2016zah, Gorham:2018ydl}. The two anomalous ANITA events (AAE) are those consistent with a steeply upgoing $\nu_{\tau}$ interpretation.} \label{tab:candidates}
	\begin{tabular}{ l |c |c |c}
	    \hline
	    \hline
	    & \textbf{AAE-061228} & \textbf{AAE-141220} & \textbf{AAC-150108} \\ \hline \hline
		Event, Flight & 3985267, ANITA-I & 15717147, ANITA-III &  83139414, ANITA-III  \\
		Detection Channel & Geomagnetic & Geomagnetic & Askaryan \\
		Date, Time (UTC) & 2006-12-28, 00:33:20 & 2014-12-20, 08:33:22.5 & 2015-01-08, 19:04:24.237 \\
		RA, Dec (J2000)\tnote{$^1$} & 282$^\circ$.14, +20$^\circ$.33 & 50$^\circ$.78, +38$^\circ$.65 & 171$^\circ$.45, +16$^\circ$.30\\
		Localization Uncertainty\tnote{$^2$} & 1$^\circ$.5 $\times$ 1$^\circ$.5, 0$^\circ$.0 & 1$^\circ$.5 $\times$ 1$^\circ$.5, 0$^\circ$.0 & 5$^\circ$.0 $\times$ 1$^\circ$.0, +73$^\circ$.7 \\
		Reconstructed Energy (EeV) & 0.6 $\pm$ 0.4 & $0.56^{+0.30}_{-0.20}$ & $\geq$ 10\\
		Earth Chord Length (km) & 5740 $\pm$ 60 & 7210 $\pm$ 55 & -
	\end{tabular}
	\begin{tablenotes}\footnotesize
    \item[$^1$] Sky coordinates are projections from event arrival angles at ANITA
    \item[$^2$] Expressed as major and minor axis standard deviations, position angle. This angle describes the rotation of the major axis relative to the north celestial pole turning positive into right ascension. 
    \end{tablenotes}
	\end{threeparttable}
	\vspace{0.05in}
\end{table*}

The interpretation of these events as extremely high energy upgoing neutrinos poses many challenges  under Standard Model assumptions. First, from the observation angles and reconstructed energies of the ANITA events, neutrinos are extremely unlikely to traverse the long chord lengths \citep{Gorham:2016zah}, even after accounting for the probability increase due to $\nu_{\tau}$ regeneration. Second, if these events are of cosmogenic origin, they would imply fluxes that are in severe tension with limits set by multiple experiments \citep{Aab:2015kma, Zas:2017xdj, Aartsen:2016ngq, Aartsen:2018vtx} as well as a self-inconsistency from ANITA data alone. For an isotropic flux of cosmogenic neutrinos, ANITA should have detected many more events at other elevation angles than those of the anomalous ANITA events (AAE) as the detector differential acceptance changes with the observation angle \citep{Romero-Wolf:2018zxt}.

On the other hand, if the origin of \textcolor{black}{the AAE} is considered to be from individual cosmic accelerators, there is no inconsistency with diffuse extremely high energy flux limits. This is especially true for accelerators with short characteristic timescales of emission, as many current limits on neutrino point sources are for integrated emission over various experiments' live times \citep{Aartsen:2018ywr} and also as the acceptance of ANITA to a specific location in the sky changes throughout the detector's flight. If we assume that ANITA detected single events of 1 EeV from a cosmic accelerator with an $E^{-\gamma}$ emission power-law spectrum, then one should expect also a larger flux of neutrinos at TeV--PeV energies, where IceCube will be sensitive. Significant correlation between IceCube and ANITA data would not only provide evidence for a neutrino point source, it would also eliminate nonastrophysical explanations of the AAE, such as background and systematics, or nonastrophysical models, which invoke physics beyond the Standard Model.

The focus of this work is to use IceCube to investigate the hypothesis that the ANITA events were from neutrino point sources, considering several neutrino emission time profiles. In section \ref{sec:icecube}, we discuss the IceCube Neutrino Observatory and the event samples used for these analyses. In section \ref{sec:methods}, we describe the analysis techniques, and we summarize the results in section \ref{sec:results}. In sections \ref{sec:discussion} and \ref{sec:conclusion}, we investigate neutrino propagation through large Earth chord lengths to discuss the implications of our results.

\section{Data Sample}
\label{sec:icecube}

IceCube is a cubic-kilometer neutrino detector with 5160 digital optical modules (DOMs) instrumented on 86 cable strings in the clear glacial ice at the geographic South Pole, at depths between 1450 m and 2450 m \citep{Achterberg:2006md, Aartsen:2016nxy}. Neutrinos are detected through the Cherenkov radiation emitted by secondary particles produced by neutrino interactions in the surrounding ice or bedrock. Each DOM consists of a 10-inch photomultiplier tube, onboard readout electronics, and a high-voltage board, all contained in a pressurized spherical glass container \citep{Abbasi:2008aa, Abbasi:2010vc}. Parameterization of the scattering and absorption of the glacial ice allows accurate energy and directional reconstruction of neutrino events~\citep{Aartsen:2013rt}. 

The improved reconstruction techniques adopted to create the event selection \citep{Carver:2019jcd, Aartsen:2019fau} include updates in the direction reconstruction \citep{Ahrens:2003fg, Aartsen:2013bfa} to use information on the deposited event energy in the detector. The median angular resolution benefits from a 10\% improvement above 10 TeV (where it is smaller than 0.60$^{\circ}$) compared to previous selections \citep{Aartsen:2016oji}.

While in the southern sky the trigger rate is dominated by atmospheric muons from cosmic-ray air showers, all of the ANITA candidates have best-fit directions in the northern sky. Here, Earth attenuates the majority of the atmospheric muon signal, and the background at final selection level in the northern sky is dominated by atmospheric muon neutrinos from cosmic-ray air showers \citep{Haack:2017dxi}. Poorly reconstructed atmospheric muons from the southern sky as well as neutrino-induced cascades are also non-negligible backgrounds in this region of the sky and are removed using a multivariate boosted decision tree trained to distinguish between neutrino-induced muon tracks, atmospheric muons, and cascades, which is described in \citep{Carver:2019jcd, Aartsen:2019fau}.

For the analyses presented here, we focus on the full detector configuration of 86 strings, spanning a time window from 2011 to 2018. Approximately 900,000 events from 2532 days are analyzed. 

\section{Likelihood analyses}
\label{sec:methods}
Many previous IceCube analyses searching for neutrino point sources relied on significant spatial clustering of IceCube data alone or of significant association with known populations of astrophysical objects \citep{Aartsen:2018ywr, Aartsen:2016tpb, Aartsen:2016oji, Aartsen:2013uuv, Abbasi:2010rd, Aartsen:2014cva}. Here, we adopt the procedure described in \citep{Schumacher:2019qdx} to search for counterparts to ANITA events. Namely, we perform three separate analyses to test different temporal hypotheses in the neutrino emission. Each of these analyses incorporates the information from the localization of the ANITA events through a joint likelihood. The sky is divided into grid positions, $\mathbf{x}_s$, and at each point we maximize the likelihood, $\mathcal{L}$, with respect to the expected number of signal events, $n_s$, and other signal parameters contained in the variable $\mathbf{\alpha}$ depending on the different signal hypotheses tested as described in sections \ref{sub:flare} and \ref{sub:steady}. This likelihood is given by

\begin{eqnarray} \label{eq:general_likelihood}
    \mathcal{L} = \lambda \prod_{i=1}^{N} \Bigg( \frac{n_s}{n_s + n_b}S(\mathbf{x}_i, \mathbf{x}_s, \mathbf{\alpha})  + \frac{n_b}{n_s + n_b}B(\mathbf{x}_i, \mathbf{x}_s) \Bigg) P_A(\mathbf{x}_s) ,
\end{eqnarray}
where $n_b$ is the expected number of observed background events and $N$ is the total number of observed events in the time window. The vector $\mathbf{x}_i$ contains the event observables such as its reconstructed energy, direction, and reconstruction uncertainty. $P_A$ is the spatial probability distribution function (PDF) of ANITA events, which are included in Table~\ref{tab:candidates}. $B$ describes the energy and declination PDF of our background, which is parameterized from data and is the same among all analyses. Temporal terms in $B$ are described in sections \ref{sub:transient} and \ref{sub:flare}. While the signal PDF $S$ describes the signal hypothesis, the parameter $\lambda$ modifies the likelihood formalism in order to take into account low-statistics problems in some of the analyses. In general, the signal PDF, $S$, is defined as
\begin{equation}
    S = S^{space}(\mathbf{x}_i, \mathbf{x}_s, \mathbf{\sigma_i})\cdot S^{energy}(E_i,\delta_i,\gamma)\cdot S^{time} \; .
\end{equation}
These three terms reflect the spatial, energy, and time PDFs, respectively, of our signal hypothesis. The spatial term, $S^{space}$, expresses the probability for an event
with best-fit reconstructed direction $\mathbf{x}_i$ to originate from a source at the direction $\mathbf{x}_s$,
according to a two-dimensional Gaussian function with angular resolution $\sigma_i$. 
The energy PDF, $S^{energy}$, 
describes the probability of obtaining an event with reconstructed energy $E_i$
given a declination $\delta_i$ under the hypothesis of an  
$E^{-\gamma}$ power-law energy spectrum, which helps differentiate signal from the known atmospheric backgrounds in our event selection. The time term, $S^{time}$, describes the time PDF of events observed from the source. 
While the spatial term is shared between all analyses, the energy and temporal terms are unique to each individual analysis. This joint likelihood procedure is carried out in three complementary search strategies: \textit{prompt}, \textit{rolling}, and \textit{steady}. 

\begin{figure*}[t!]
\centering
\vspace{0.0cm}
    \includegraphics[width=0.99\textwidth=]{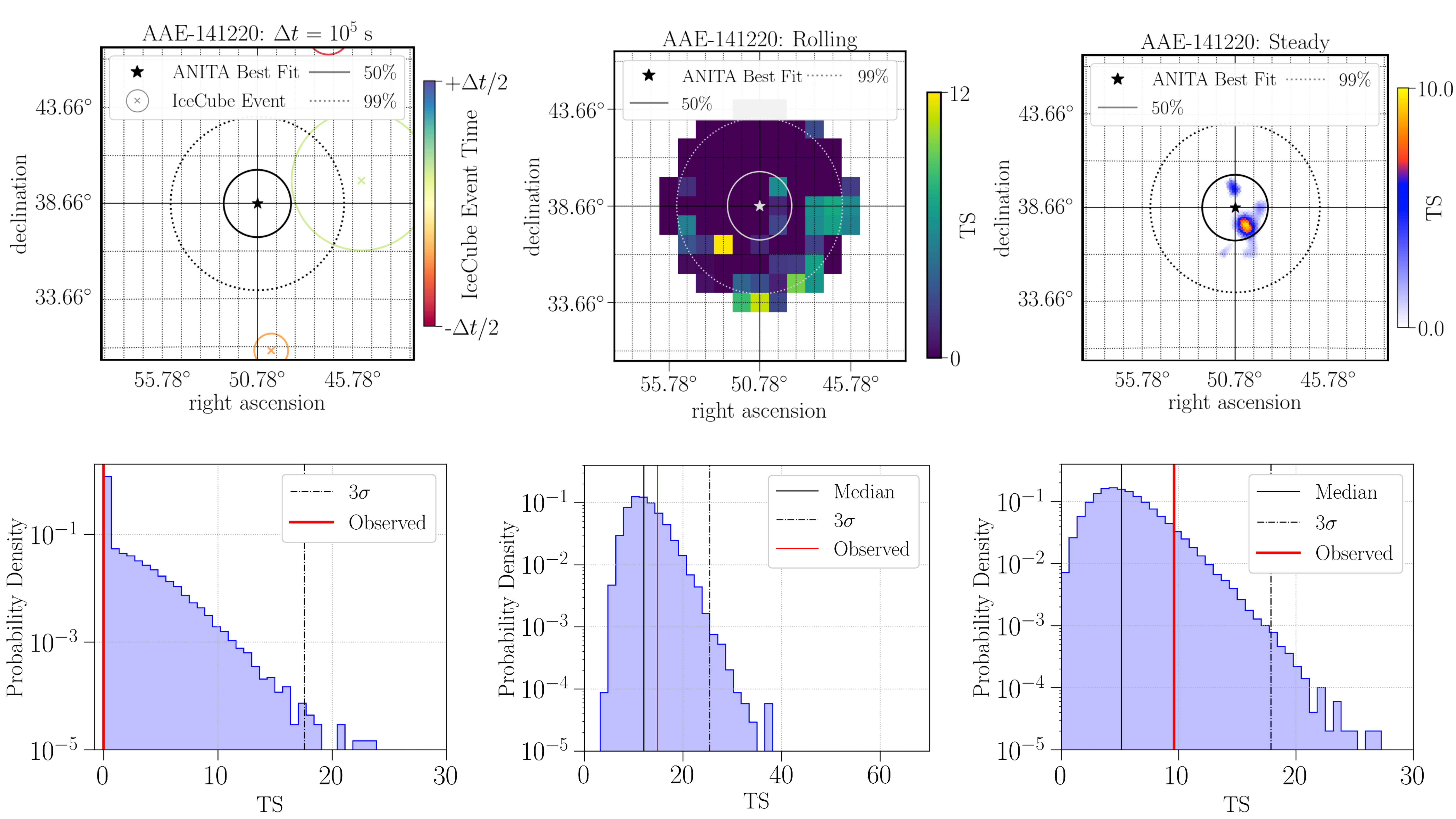}
\caption{Sky maps (top) and TS distributions (bottom) for AAE-141220 for the prompt (left), rolling (middle), and steady (right) analyses. Observed TS values (shown in red) are compared to distributions from time-scrambled data realizations to quantify the significance. \textcolor{black}{In all sky maps, solid (dotted) lines represent 50\% (99\%) containment of the reconstructed direction of the events.} In the prompt analysis sky map, the best-fit location of each IceCube event is represented with an \texttt{x}, and the size of the circle represents the uncertainty (50\% containment) on the event's reconstruction, with color representing the IceCube event arrival time relative to the ANITA event. \textcolor{black}{Both the sky map and TS distribution for this analysis are for the $10^5$ s time window}. In the rolling and steady analysis sky maps, color reflects the TS values defined in sections \ref{sub:flare} and \ref{sub:steady} respectively.} \label{fig:skymaps}
\end{figure*}

\subsection{Prompt}
\label{sub:transient}
The first analysis searches for IceCube events in spatial coincidence with the ANITA events in short time windows, $\Delta t$, centered on each ANITA event. We call this period the \textit{on-time window}. This is equivalent to setting $S^{time}$ equal to a uniform PDF in this on-time window and to zero for all times outside this window. To help distinguish potential signals for time windows in which the expected number of background events is small, we set \begin{equation}
    \lambda = \frac{(n_s + n_b)^N}{N!}\cdot {\rm e} ^{-(n_s + n_b)}
\end{equation} 
as in \citep{Aartsen:2017zvw, Aartsen:2014aqy}. Due to the small statistics for short time windows, the likelihood is only maximized with respect to $n_s$, and the energy dependence in $S^{energy}$ is fixed to an $E^{-2}$ spectrum. To account for the temperature dependence of atmospheric muon rates \citep{Aartsen:2013jla}, we determine $n_b$ by calculating the rate of events from the surrounding five days of data on either side of our on-time window. Taking the logarithmic likelihood ratio between the maximum likelihood and that of the null hypothesis results in our test statistic (TS), defined as
\begin{equation} \label{eq:transient_TS}
    \text{TS} = - 2\hat{n_s} +  \sum_{i=1}^{N}2\log \left[ 1 + \frac{\hat{n_s} S(\mathbf{x}_i, \mathbf{x}_s)}{n_b B(\mathbf{x}_i)} \right] + 2 \log \left[ \frac{P_A (\mathbf{x}_s)}{P_A (\mathbf{x}_0)} \right],
\end{equation}
where $\mathbf{x}_0$ is the reported best-fit location of the ANITA event \textcolor{black}{and $\hat{n}_s$ is the value of $n_s$ that maximizes the likelihood}. TS is calculated for all $\mathbf{x}_s$, and the maximum value is reported. For this analysis, $P_A$ is a two-dimensional Gaussian assuming the localization uncertainties reported in Table~\ref{tab:candidates}. As we are not motivated by a specific astrophysical class of objects with characteristic timescales of emission, 
we consider constant emission over various time windows for each of the ANITA events. This technique is similar to previous IceCube searches for gamma-ray bursts and fast radio bursts \citep{Aartsen:2017zvw, Aartsen:2014aqy}. AAE-061228 is excluded from this analysis because it occurred before IceCube had attained a full detector configuration. For AAC-150109 we consider three separate time windows: 10 s, $10^3$ s, and $10^5$ s. During the event time of AAE-141220, IceCube was temporarily not collecting data, due to a run transition that had begun approximately 0.5 seconds before the event and lasting for about one minute. Because of this, we only investigate hypotheses of constant emission over two time windows ($10^3$ s and $10^5$ s), where the period of time from the run transition is not an appreciable portion of our on-time window.

\subsection{Rolling}
\label{sub:flare}
The second analysis also searches for temporal and spatial clustering of IceCube events but does not require the temporal coincidence between IceCube and ANITA events. In this untriggered analysis~\citep{Braun:2009wp,Aartsen:2015wto}, we assume a Gaussian time dependence to parameterize
a limited duration increase in the emission of the source:
\begin{eqnarray} \label{eq:SigPDFUntrig}
    S^{time} = \frac{1}{\sqrt{2\pi}\sigma}e^{-\frac{(t_0 - t_i)^2}{2\sigma_t^2}},
\end{eqnarray}
where $t_0$ and $\sigma_t$ are the Gaussian mean time and Gaussian width of the
flare, respectively. In the limit of large $N$, we are free to set $\lambda$ to 1, and the increase in statistics allows us to fit for $\gamma$ in the range $1 \leq \gamma \leq 4$ in addition to $n_s$, as is done in many previous IceCube analyses \citep{Aartsen:2016tpb, Aartsen:2016oji, Aartsen:2013uuv, Abbasi:2010rd, Aartsen:2014cva}. Additionally, we set $n_s + n_b$ to be equal to the number of events, $N$. 
The TS for this analysis is then
\begin{eqnarray} \label{eq:TSUntrig}
    \text{TS} = -2 \, \text{log} \Bigg[\frac{\text{T}}{\sqrt{2\pi}\hat{\sigma_t}} \times \frac{\mathcal { L } (n_s=0)}{\mathcal{L}(\hat{n_s},\hat{\gamma},\hat{\sigma_t},\hat{t_0})}\Bigg],
\end{eqnarray}
where $\hat{n_s},\hat{\gamma},\hat{\sigma_t},\hat{t_0}$ are the best-fit values
from the likelihood maximization and $T$ is the total live time of the data-taking period. 
The multiplicative factor in front of the likelihood ratio in Eq.~\ref{eq:TSUntrig} 
is a marginalization term to avoid undesired biases toward finding short flares, as explained in \citep{2010APh....33..175B}.
The TS is calculated at the positions of a coarse sky grid (1$^{\circ} \times 1^{\circ}$ bin widths), built at the central 
coordinates of the ANITA events and covering 99.9\% of the their 
two-dimensional spatial PDFs, but sets $P_A$ to be a uniform distribution covering this extended region. As the PDF is taken to be uniform in this analysis, there is no term in the TS that is dependent on $P_A$.
The location of the maximum TS from the coarse search is then used
as a seed to perform a further likelihood maximization, where the direction of the source, $\mathbf{x}_s$, is also reconstructed. 

\subsection{Steady}
\label{sub:steady}
The third and final analysis tests for spatial clustering over seven years of IceCube data, assuming constant emission in the signal hypothesis, by setting $S^{time}$ to be a uniform PDF over the entire data collection period. As in the rolling analysis, we take $\lambda$ to be 1 and fit for $\gamma$ in the likelihood maximization process. At all $\mathbf{x}_s$ we calculate the redefined TS
\begin{equation}
    \text{TS} = 2 \cdot \log \left[ \frac { \mathcal { L } \left( \mathbf { x } _ { s } , \hat { n } _ { s } , \hat { \gamma } \right) } { \mathcal { L } \left( \mathbf { x } _ { s } , n _ { s } = 0 \right) } \right] + 2 \log \left[ \frac{P_A (\mathbf{x}_s)}{P_A (\mathbf{x}_0)} \right]\; , \qquad 
\end{equation}
with best-fit values $\hat{n}_s$ and $\hat{\gamma}$. The PDF of ANITA events in this analysis is taken to be the same as in the prompt analysis, namely, a two-dimensional Gaussian.

\section{Results}
\label{sec:results}

No significant correlation is found in any of the analyses above the expectation from background. In order to calculate p-values, results are compared against pseudo-experiments from time-scrambled data \citep{Aartsen:2015wto}. The most significant observation results from the steady search for AAE-141220, with a p-value of 0.08 \textcolor{black}{before trials correction.}

\begin{figure*}
    \centering
    \includegraphics[width=0.99\textwidth]{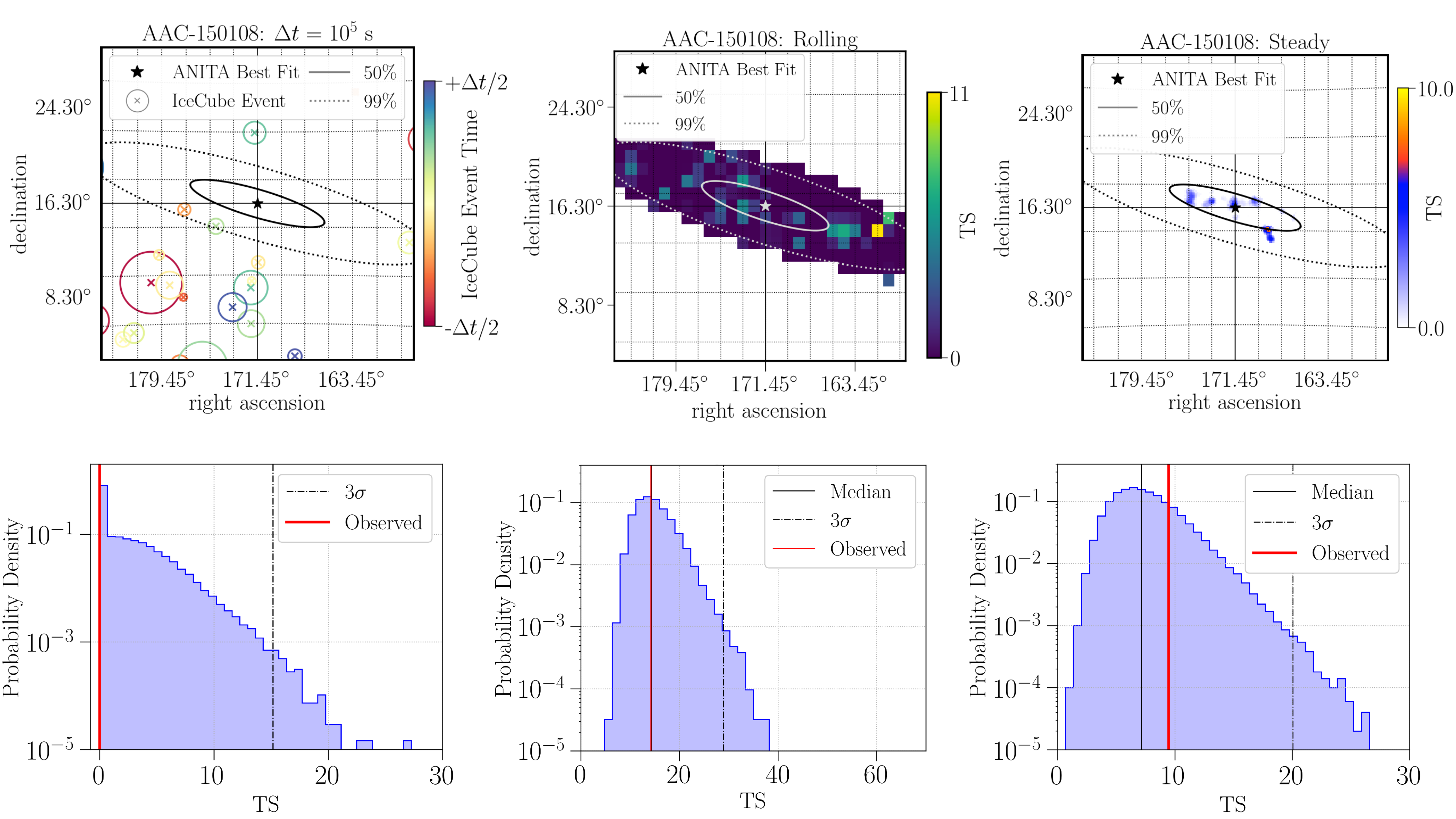}
    \includegraphics[width=0.99\textwidth]{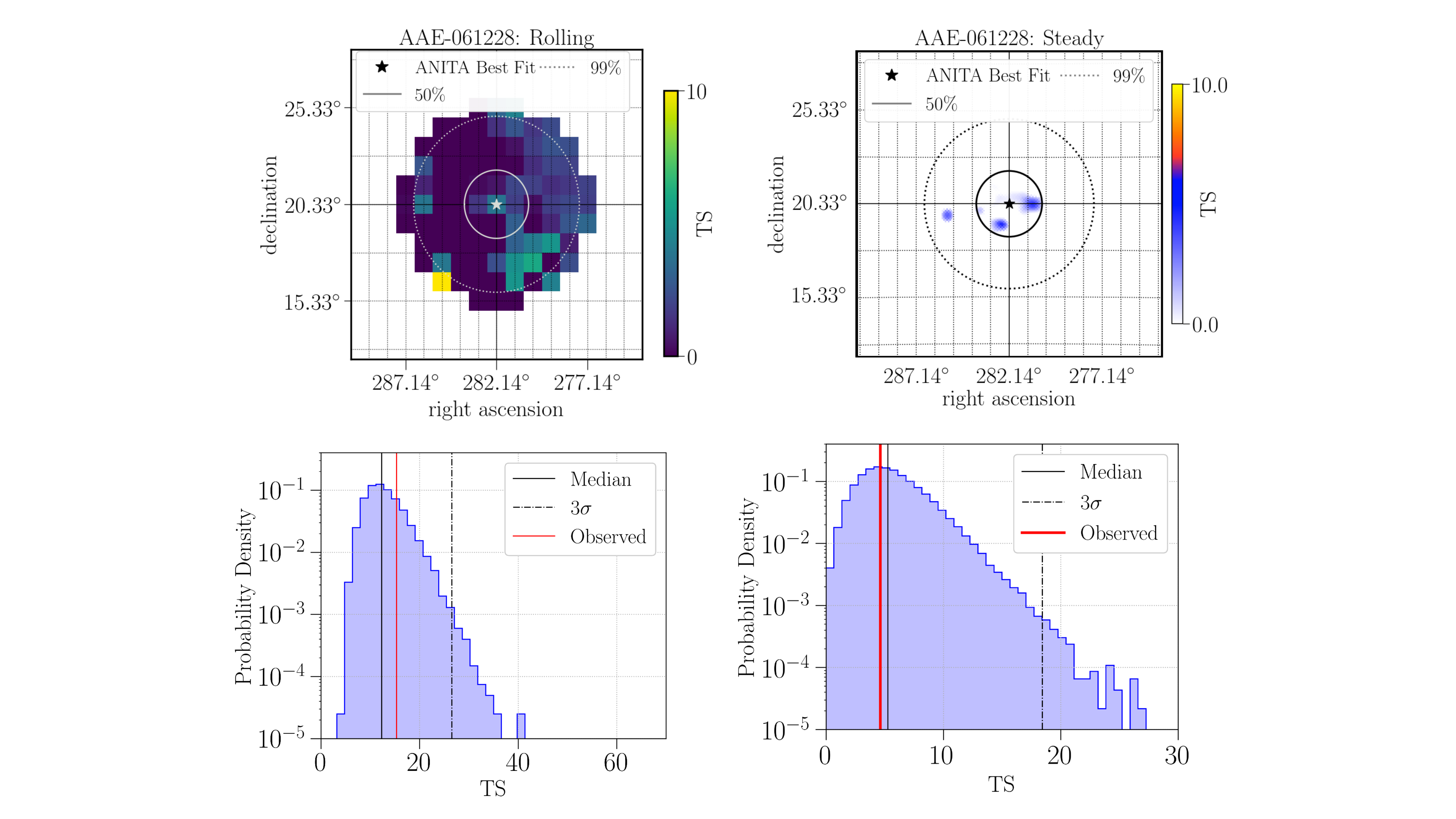}
    \caption{(Top two rows) Skymaps and TS distributions from all three analyses for AAC-150108. For AAE-061228, IceCube was not in a full detector configuration at the time of the event, and thus only the steady and rolling analyses were used to search for neutrino emission. Skymaps and TS distributions for these analyses are displayed in the bottom two rows.} \label{fig:skymaps_extra}
\end{figure*}

Figure \ref{fig:skymaps} displays the sky maps for the prompt, rolling, and steady analyses from left to right in the top panels for AAE-141220. Bottom panels of Figure \ref{fig:skymaps} show the comparison of the observed TS values for each analysis, at the position of the red lines, to their respective TS distributions from pseudo-experiments using time-scrambled data. Similar plots for AAE-061228 and AAC-150108 are displayed in Figure~\ref{fig:skymaps_extra}.

\begin{figure}
\vspace{0.05in}
\centering
  \includegraphics[width=0.47\textwidth]{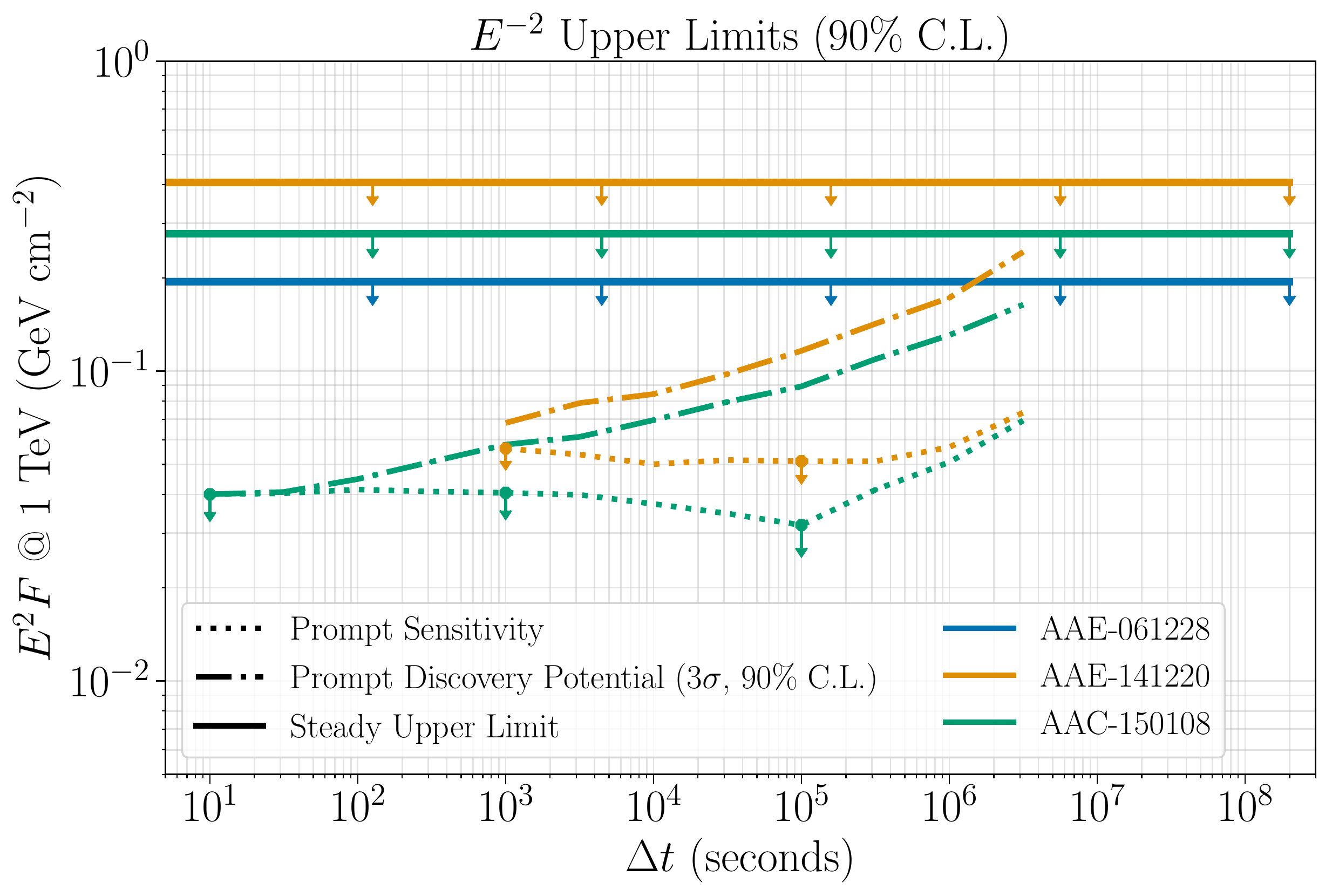}
  \label{fig:upper_limits}
  \caption{Sensitivity (dotted) and upper limits (arrows) (90\% confidence level) on the time-integrated $\nu_{\mu} + \bar{\nu}_{\mu}$ flux normalization for an $E^{-2}$ source spectrum as a function of $\Delta t$ from the prompt analysis, compared to the upper limits (solid) from the steady analysis. The central 90\% intervals of the expected neutrino energies for these spectra are 1 TeV-1 PeV. \textcolor{black}{For the prompt analysis, we also include the discovery potential, which is the flux that results in a 3$\sigma$ result, pre-trials, in 90\% of pseudo-experiments.}}
\end{figure}

In the absence of a significant signal, upper limits (90\% confidence level) for the time-integrated $\nu_{\mu} + \bar{\nu}_{\mu}$ flux are set for each ANITA event where possible using the prompt and steady analyses (Figure \ref{fig:upper_limits}). To calculate upper limits, locations are sampled according to the per-event PDFs reported by ANITA, injecting the same level of flux at each sampled location, and running each iteration through the full analysis procedure, which maximizes the joint likelihood at all locations in the sky. This allows us to place upper limits on point sources whose locations are distributed according to the per-event PDF reported by ANITA. We set these limits for an assumed spectrum given by
\begin{equation}
    \Phi (E, t) = \frac{\diff N_{\nu_{\mu} + \bar{\nu}_{\mu}}}{\diff E \diff A \diff t} =  \Phi_0 \Big(\frac{E}{E_0}\Big)^{-2} \; , 
\end{equation}
where $\Phi_0$ is a normalization constant on a point-source flux, which carries units of $\rm{GeV}^{-1}\rm{cm}^{-2}\rm{s}^{-1}$. We constrain the time-integrated muon neutrino flux, $E^2F$, where 
\begin{equation}
    E^2 F = E^2 \int \Phi(E, t) \diff t \; .
\end{equation} 
All of the limits we calculate are provided in Table~\ref{tab:results}. In the case that an upper limit fluctuates below the sensitivity, we conservatively set the upper limit to the sensitivity value. Prompt limits are placed at the specified time windows for emission centered on the ANITA event times, whereas limits from the steady analysis are for emission over the live time of our data sample. This hard spectrum was chosen conservatively because with the observation of EeV events by ANITA, if the underlying spectrum is softer, then the expected number of observable neutrinos for IceCube would increase. As the time-integrated flux sensitivity for the triggered analysis begins to worsen past $10^5$ s, upper limits for $\Delta t > 10^5$ s are only set using the time-integrated approach. 

\begin{table*}
    \caption{Analysis results and upper limits.  Upper limits (90\% C.L) are on the time-integrated $\nu_{\mu} + \bar{\nu}_{\mu}$ power law flux ($E^{-2}$ ) from a point source following the spatial probability distribution provided by ANITA. Limits are set assuming constant emission over a fixed time window. As the temporal profile of emission is fit in the rolling analysis, no upper limits are placed from that analysis. Time windows for the steady and rolling analyses are listed as the IceCube seasons analyzed, where IC86-I contains 2.88$\times 10^7$ s of data and IC86-II--IC86-VII contain 1.90$\times 10^8$ s. \textcolor{black}{All $p$-values are not trial-corrected for the number of searches considered.}}
    \centering
    \begin{tabular}{l | c| c| c | c} \hline
     Event & Analysis & Time Window  & $p$-value & Upper limit (GeV $\cdot$ cm$^{-2}$)  \\ \hline \hline
     \multirow{3}{*}{AAE-061228} & Steady & IC86-I - IC86-VII & 0.606 &  0.195 \\ \cline{2-5}
     & \multirow{2}{*}{Rolling} & IC86-I & 0.562 & - \\ 
     & & IC86-II - IC86-VII & 0.208 & - \\ \hline
     \multirow{5}{*}{AAE-141220} & & 10 s & - & - \\ 
	 & Prompt & $10^3$ s & 1.0 & 0.053 \\ 
	 & & $10^5$ s & 1.0 & 0.051 \\ \cline{2-5}
	 & Steady & IC86-I - IC86-VII & 0.081 & 0.401 \\ \cline{2-5}
	 & \multirow{2}{*}{Rolling} & IC86-I & 0.342 & - \\ 
     & & IC86-II - IC86-VII & 0.224 & - \\ \hline
	 \multirow{5}{*}{AAC-150108} & & 10 s & 1.0 & 0.040 \\ 
	 & Prompt & $10^3$ s & 1.0 & 0.041 \\ 
	 & & $10^5$ s & 1.0 & 0.032 \\ \cline{2-5}
	 & Steady & IC86-I - IC86-VII & 0.210 & 0.278 \\ \cline{2-5}
	 & \multirow{2}{*}{Rolling} & IC86-I & 0.636 & - \\ 
     & & IC86-II - IC86-VII & 0.512 & - \\ \hline
    \end{tabular}
    \label{tab:results}
    \vspace{0.2in}
\end{table*}
\section{Discussion}
\label{sec:discussion}
For many astrophysical sources, power-law spectra in photons are common over finite energy ranges. Additionally, diffusive shock acceleration models suggest that the neutrino spectrum, as well as gamma rays from pion decay, should follow a power-law spectrum, justifying the choice of testing power laws for corresponding neutrino spectra. However, for the \textcolor{black}{AAE}, interpolating a power law between the energy range at which IceCube is sensitive to the best-fit ANITA event energies could pose a problem. For soft spectra, events detected by ANITA would suggest that many events would be detectable at IceCube. For hard spectra, extrapolating between IceCube and ANITA would imply dramatic bolometric neutrino luminosities for any point source.

However even in the case of non-power-law neutrino emission, the limits we can set on muon neutrinos in the TeV--PeV energy range can constrain generic fluxes of incident tau neutrinos with EeV energies. As has been shown in \citep{Safa:2019ege}, any incident flux with an EeV $\nu_{\tau}$ component that traverses large Earth chord lengths will result in a secondary flux of lower energy neutrinos, to which IceCube would be sensitive. We use the same prescription here to analyze how constraining our limits are on a generic point source flux that includes EeV neutrinos.

For any incident flux of neutrinos from the northern sky, $\Phi(E_{\nu}, t)$, the number of expected detected tau-neutrino--induced muon events at IceCube is given by 

\begin{figure}
    \centering
    \includegraphics[width=0.45\textwidth]{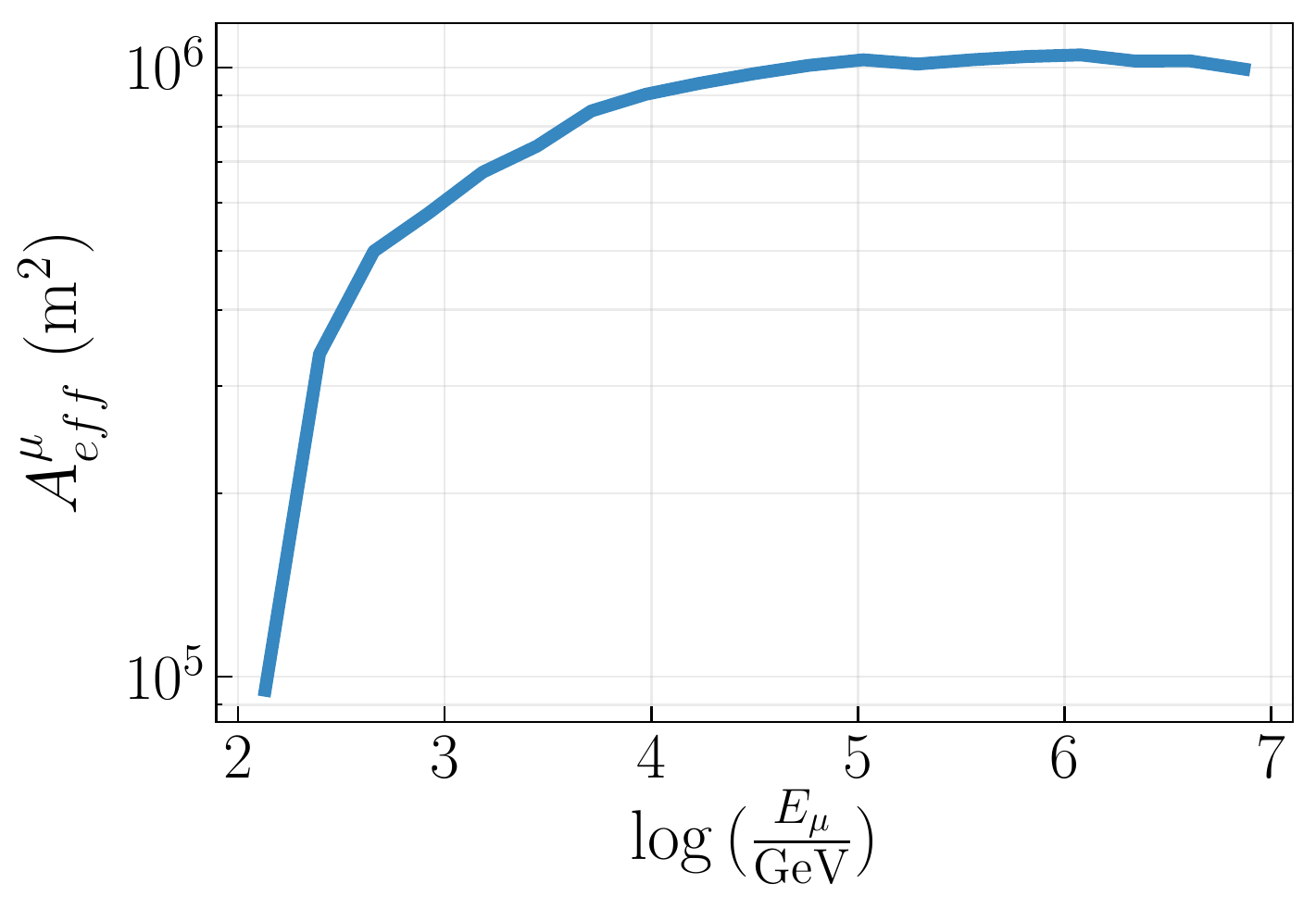}
    \caption{Effective area of the IceCube event selection to muons from the northern sky, incident on a volume 1.5 km away from the edge of the detector. $E_{\mu}$ is the muon energy incident on this volume.}
    \label{fig:effective_area}
\end{figure}

\begin{widetext}
\begin{align}
\langle N _ { \text {IceCube} } ^ { \mu } \rangle & = \int \diff E _ { \mu } \int \diff E _ { \tau } \int \diff E _ { \nu } \Phi \left( E _ { \nu }, t \right) P _ { \tau } ^ { s u r v } \left( E _ { \nu } \right) \frac { \diff N _ { \tau } \left( E _ { \tau } \right) }{ \diff E _ { \tau } }  \frac{\Gamma_{\tau\rightarrow\mu}}{\Gamma_{\rm{total}}} \frac { \diff N _ { \mu } } { \diff E _ { \mu } } \left( E _ { \tau } , E _ { \mu } \right) A _ { e f f } ^ { \mu } \left( E _ { \mu } \right) \Delta T \\ & + \int \diff E _ { \mu } \int \diff E _ { \tau } \int \diff E _ { \nu } ^ { \prime } \int \diff E _ { \nu } \Phi \left( E _ { \nu }, t \right) P _ { \nu } \left( E _ { \nu } , E _ { \nu } ^ { \prime } \right) \frac { \diff N _ { \nu } } { \diff E _ { \nu } ^ { \prime } } \left( E _ { \nu } ^ { \prime } \right)  N^{p} \left( E^{\prime}_{\nu} \right) \frac { \diff N _ { \tau } } { \diff E _ { \tau } } \left( E ^{\prime} _ { \nu} ; E _ { \tau} \right)  \frac{\Gamma_{\tau\rightarrow\mu}}{\Gamma_{\rm{total}}} \frac { \diff N _ { \mu } } { \diff E _ { \mu } } \left( E _ { \tau} ; E _ { \mu} \right) A _ { e f f } ^ { \mu } \left( E _ { \mu } \right) \Delta T \; , \nonumber
\end{align}
\end{widetext}
where the first contribution is from emerging tau-leptons that would decay to muons and then pass an IceCube event selection. The second contribution is from the remaining $\nu_{\tau}$ flux, the majority of which has cascaded down in energy. $N^{p} (E_{\nu})$ is the number of targets effectively seen by an incident neutrino with energy $E_{\nu}$. The effective area of this event selection to muons incident on the detector is displayed in Figure~\ref{fig:effective_area}. $P _ { \tau } ^ { s u r v } \left( E _ { \nu } \right)$ and $P _ { \nu }(E _ { \nu })$ represent the survival probability of a $\tau$-lepton and $\nu_{\tau}$, given an incident neutrino energy, respectively, and $\Gamma_{\tau\rightarrow\mu}\big/\Gamma_{\rm{total}}$ represents the branching ratio for the tau-decay to muon channel, which is approximately 18\%. 

Similarly, for ANITA, the number of expected events from upgoing $\tau$-leptons is given by 

\begin{equation}
\begin{aligned}
    \langle N ^ { \tau }_{\mathrm{ANITA}} \rangle =& \iint \Big( \diff E _ { \nu } \diff E^{\prime} _ { \nu } \Phi \left( E _ { \nu }, t \right) \frac { \diff N \left( E^{\prime} _ { \nu } \right) } { \diff E^{\prime} _ { \nu } }  \\ & \qquad \times \xi _ { a c c } \left( E^{\prime} _ { \nu } \right) \Delta T \Big) ,
\end{aligned}
\end{equation}
where $\xi_{acc}$ represents ANITA's acceptance to $\tau$-lepton air showers, taken from \citep{Romero-Wolf:2018zxt}. \textcolor{black}{Values for the acceptance at angles that would require an incident neutrino to traverse a large column depth are set to the acceptance near the horizon. We take the value at an angle corresponding to the maximum acceptance before absorption effects dominate.} This removes absorption effects in the reported acceptance, which is accounted for separately with the code used to propagate these fluxes, \texttt{TauRunner}, described in \citep{Safa:2019ege, Safa:2019icrc}. We focus our analysis on the nonobservation of coincident events in IceCube at $\Delta T = 10^3$~s. A similar procedure can be applied to longer time windows. Qualitatively, it would result in similar limits up to the lifetime of the ANITA flight. For longer emission timescales, limits from IceCube become even more constraining as the implied normalization on the ANITA flux would have to increase to compensate for the fraction of time during which ANITA was not taking data.

\begin{figure}
    \centering
    \includegraphics[width=0.42\textwidth]{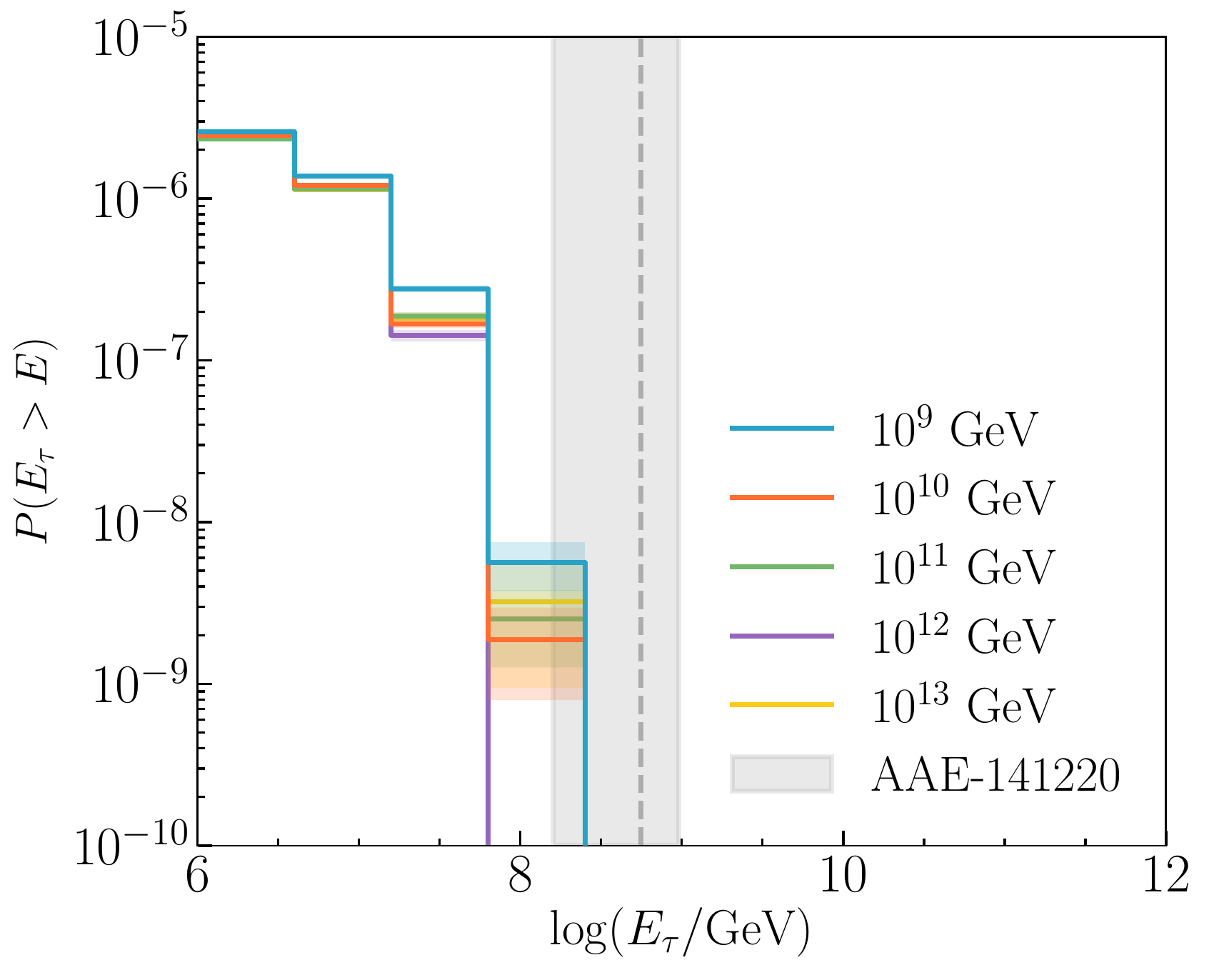}
    \caption{Normalized cumulative distributions for Earth-emerging tau-leptons. Colors correspond to the incoming tau-neutrino energy, and the gray band is the 95\% containment on the error of the reconstructed shower energy of AAE-141220.}
    \label{fig:tau_exit}
\end{figure}

To make as conservative a statement as possible, we inject fluxes described by delta functions in energy,
$\Phi(E_{\nu}, t)= \Phi_0 \delta(E_{\nu} - E_0)$, where now the normalization carries units of $\rm{cm}^{-2}\rm{s}^{-1}$. After propagating these mono-energetic fluxes, we record what fraction of the incident flux results in a detectable signal at ANITA. We repeat this procedure for a variety of injected initial neutrino energies so that we can find the energy that yields the maximum probability of a $\tau$-lepton arriving at ANITA with an energy within the quoted reconstructed energy bounds. We find that the optimal flux for ANITA corresponds to an injected $\nu_{\tau}$ flux with $E_0 = 1$ EeV. Normalized cumulative distributions from secondary $\tau$-leptons are shown in Figure~\ref{fig:tau_exit} for injected neutrinos at angles corresponding to the best-fit reconstructed direction of AAE-141220.

\begin{figure*}
\centering
  \includegraphics[width=0.65\textwidth]{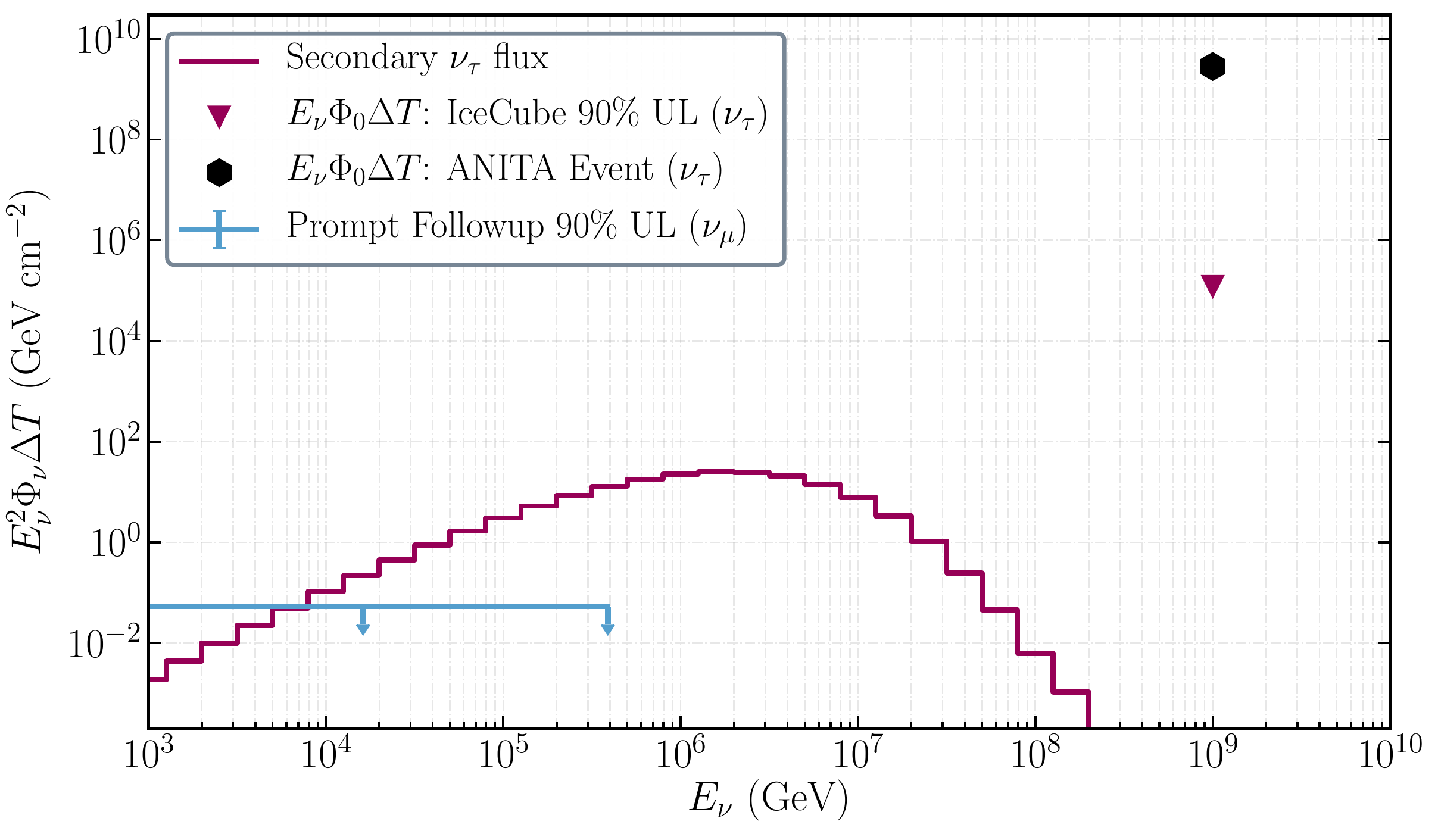}
  \label{fig:taurunner_limits}
  \caption{Upper limits (90\% C.L.) placed by calculating the secondary neutrino flux (purple histogram) from an incident flux of EeV neutrinos assuming constant emission over $10^3$ s and comparing to the nonobservation of IceCube events in the prompt analysis described in Sect.~\ref{sub:transient} for AAE-141220. The flux implied by the ANITA observations (black), represented in this figure as $E_{\nu}\Phi_0 \Delta T = E_{\nu} \Delta T \int \Phi(E_{\nu}, t) \diff E_{\nu}$, using information about ANITA's acceptance \citep{Romero-Wolf:2018zxt} overshoots this upper limit (purple arrow) by many orders of magnitude. For comparison, upper limits on the time-integrated muon-neutrino flux from the prompt analysis are shown in blue. All fluxes are per flavor $\nu + \bar{\nu}$.}
\end{figure*}

We next inject a flux of EeV tau neutrinos and find the spectral shape of the secondary $\nu_{\tau}$ flux that would be incident on IceCube. As we observed zero coincident events in the time window of $10^3$ s around AAE-141220 in the prompt analysis, we calculate the maximum allowed flux normalization (at 90\% confidence level) on the primary flux that would evade this nonobservation. 
The results are displayed in Figure~\ref{fig:taurunner_limits}. 

Although IceCube's sensitivity peaks many orders of magnitude below the reconstructed energies of the ANITA events, the limits set on any potential neutrino source that created AAE-141220 are more constraining by several orders of magnitude than the implied flux by the ANITA observations. If one considers constant emission over the entire live time of the IceCube event selection, then the time-integrated flux limit set by the IceCube nonobservation of AAE-141220 becomes around one order of magnitude less constraining, as is apparent in the steady limits in Figure~\ref{fig:upper_limits}. However, for the implied normalization placed by ANITA observations, this value would increase by approximately two orders of magnitude, due to the limited live time of the ANITA flight. This has the overall effect of increasing the tension between these two normalizations by approximately one more order of magnitude than for the 10$^3$ s followup shown in Figure~\ref{fig:taurunner_limits}. \textcolor{black}{It is worth noting that the logic for scaling time-integrated limits also applies to AAE-061228, even though we cannot constrain the shorter timescales for this event. However, the emergence angle of this event at ANITA was shallower than that of AAE-141220, which increases the probability of observing such an event at ANITA by approximately one order of magnitude \citep{Fox:2018syq} for the same assumed initial flux, and thus the limit on assumed long timescale emission would be about one order of magnitude less constraining than the case of AAE-141220.}

\textcolor{black}{If the intrinsic spectrum were to contain contributions from energies below 1 EeV, such as the power-law spectra tested in the analyses presented in Section~\ref{sec:methods}, this would introduce a component to which IceCube might be sensitive but which could not produce events at ANITA consistent with the AAE, and thus this additional component would strengthen the constraints displayed in Figure~\ref{fig:taurunner_limits}. Additionally, if the spectrum consisted of neutrinos of energy greater than 1 EeV, the secondary $\nu_{\tau}$ spectrum would have a similar shape to that shown in Figure~\ref{fig:taurunner_limits}, as discussed in \citep{Safa:2019ege}, and therefore the limits on the flux normalization would be constant for fluxes of higher energy, while the energy required to produce such a flux would scale with the injected energy. For that reason, these limits are conservative, and severely constrain any incident spectrum which could produce an observable event at ANITA consistent with the AAE.}

\section{Conclusion}
\label{sec:conclusion}
Recent detections of \textcolor{black}{the AAE} are considered anomalous due to the small survival probability of EeV tau neutrinos through long chord lengths. The events are known to be inconsistent with a cosmogenic interpretation but could have been produced by cosmic accelerators, specifically those with short characteristic timescales. We show here that for timescales as small as $10^3$ s, assuming AAE-141220 as originating from a neutrino source, limits set using IceCube data are in tension with the point source flux required to detect one event at ANITA by more than four orders of magnitude. These limits are constraining for a variety of flux models, from simple power laws to any generic model that includes a component at or above EeV energies. In addition to the anomalous events, we also find no evidence for a neutrino source in the direction of the neutrino candidate event from a search for Askaryan emission during ANITA-III\textcolor{black}{. As searches for Askaryan emission with ANITA have targeted a diffuse UHE cosmic neutrino flux \citep{Allison:2018cxu} and not localized point source fluxes, studies that quantify the acceptance of the ANITA detector \citep{Cremonesi:2019zzc} focus on diffuse acceptances and not effective areas for neutrino fluxes from fixed locations in the sky. For this reason, we do not provide a comparison between the limits we set here and potential implications for point source fluxes based on the observation of the AAC. With knowledge of the effective area of ANITA in the direction of the AAC, and assuming that any astrophysical flux was roughly equal in flavor upon reaching Earth, the same secondary $\nu_{\tau}$ analysis could be performed for the AAC. However, constraints from such a search would be considerably weaker than those for the AAE, as the AAC was Earth-skimming, and thus a greater fraction of any high-energy incident flux would be able to reach the ANITA detector prior to interacting deep within Earth. Therefore, this method of using secondary $\nu_{\tau}$ fluxes from UHE neutrinos in IceCube could be beneficial for future correlation searches with radio detectors and future Cherenkov detectors such as POEMMA \citep{Venters:2019xwi}.}

These new limits, in conjunction with the inconsistency of isotropic flux interpretations, leave no room for an astrophysical interpretation of the AAE in the context of the Standard Model \textcolor{black}{for time windows as short as $10^3$ s}. However, it has been shown that these events can be explained using physics beyond the Standard Model, as many models suggest that the AAE lend support for axionic dark matter, sterile neutrinos, supersymmetry, or heavy dark matter \citep{Cherry:2018rxj, Anchordoqui:2018ucj, Huang:2018als, Dudas:2018npp, Connolly:2018ewv, Fox:2018syq, Collins:2018jpg, Esteban:2019hcm, Heurtier:2019git, Heurtier:2019rkz, Abdullah:2019ofw, Anchordoqui:2018ssd, Borah:2019ciw, Chipman:2019vjm, Cline:2019snp, Esmaili:2019pcy, Hooper:2019ytr, Chauhan:2018lnq}. \textcolor{black}{Many of these models, excluding the axionic dark matter explanation \citep{Esteban:2019hcm} or those heavy dark matter scenarios that are tuned to prevent signatures in IceCube \citep{Hooper:2019ytr}, can be constrained by this nonobservation at IceCube. Dedicated tests to quantify these constraints are beyond the scope of this work and may be the focus of a future study.}
In addition to explanations that point to new physics, it has recently been suggested that the AAE could be explained by downgoing CR-induced EAS that reflected off of subsurface features in the Antarctic ice \citep{Shoemaker:2019xlt}. Another possible explanation could be coherent transition radiation from the geomagnetically induced air shower current, which could mimic an upgoing air shower \citep{deVries:2019gzs, Motloch:2016yic}. Explaining these anomalous events with systematic effects or confirming the need for new physics requires a deeper understanding of ANITA's detection volume. Efforts such as the HiCal radio frequency pulser, which has flown alongside ANITA in the last two flights \citep{Prohira:2018mmv}, are already underway to try to characterize the various properties of the Antarctic ice surface.

\section*{Acknowledgements} The IceCube Collaboration acknowledges the significant contributions to this manuscript from Anastasia Barbano, Alex Pizzuto, and Ibrahim Safa. The authors gratefully acknowledge the support from the following agencies and institutions: 
USA {\textendash} U.S. National Science Foundation-Office of Polar Programs,
U.S. National Science Foundation-Physics Division,
Wisconsin Alumni Research Foundation,
Center for High Throughput Computing (CHTC) at the University of Wisconsin--Madison,
Open Science Grid (OSG),
Extreme Science and Engineering Discovery Environment (XSEDE),
U.S. Department of Energy-National Energy Research Scientific Computing Center,
the particle astrophysics research computing center at the University of Maryland,
Institute for Cyber-Enabled Research at Michigan State University,
and the astroparticle physics computational facility at Marquette University;
Belgium {\textendash} Funds for Scientific Research (FRS-FNRS and FWO),
FWO Odysseus and Big Science programmes,
and Belgian Federal Science Policy Office (Belspo);
Germany {\textendash} Bundesministerium f{\"u}r Bildung und Forschung (BMBF),
Deutsche Forschungsgemeinschaft (DFG),
Helmholtz Alliance for Astroparticle Physics (HAP),
Initiative and Networking Fund of the Helmholtz Association,
Deutsches Elektronen Synchrotron (DESY),
and High Performance Computing cluster of the RWTH Aachen;
Sweden {\textendash} Swedish Research Council,
Swedish Polar Research Secretariat,
Swedish National Infrastructure for Computing (SNIC),
and Knut and Alice Wallenberg Foundation;
Australia {\textendash} Australian Research Council;
Canada {\textendash} Natural Sciences and Engineering Research Council of Canada,
Calcul Qu{\'e}bec, Compute Ontario, Canada Foundation for Innovation, WestGrid, and Compute Canada;
Denmark {\textendash} Villum Fonden, Danish National Research Foundation (DNRF), Carlsberg Foundation;
New Zealand {\textendash} Marsden Fund;
Japan {\textendash} Japan Society for Promotion of Science (JSPS)
and Institute for Global Prominent Research (IGPR) of Chiba University;
Korea {\textendash} National Research Foundation of Korea (NRF);
Switzerland {\textendash} Swiss National Science Foundation (SNSF);
United Kingdom {\textendash} Department of Physics, University of Oxford.

\newpage
\bibliographystyle{yahapj}
\bibliography{references}

\begin{thebibliography}{}
\providecommand\natexlab[1]{#1}
\providecommand\JournalTitle[1]{#1}

\bibitem[{Aab {et~al.}(2015)}]{Aab:2015kma}
Aab, A., {et~al.} 2015,
  \href{http://dx.doi.org/10.1103/PhysRevD.91.092008}{\JournalTitle{Phys.
  Rev.}, D91, 092008}

\bibitem[{Aartsen {et~al.}(2013{\natexlab{a}})}]{Aartsen:2013jdh}
Aartsen, M.~G., {et~al.} 2013{\natexlab{a}},
  \href{http://dx.doi.org/10.1126/science.1242856}{\JournalTitle{Science}, 342,
  1242856}

\bibitem[{Aartsen {et~al.}(2013{\natexlab{b}})}]{Aartsen:2013rt}
---. 2013{\natexlab{b}},
  \href{http://dx.doi.org/10.1016/j.nima.2013.01.054}{\JournalTitle{Nucl.
  Instrum. Meth.}, A711, 73}

\bibitem[{Aartsen {et~al.}(2013{\natexlab{c}})}]{Aartsen:2013uuv}
---. 2013{\natexlab{c}},
  \href{http://dx.doi.org/10.1088/0004-637X/779/2/132}{\JournalTitle{Astrophys.
  J.}, 779, 132}

\bibitem[{Aartsen {et~al.}(2013{\natexlab{d}})}]{Aartsen:2013jla}
Aartsen, M.~G., {et~al.} 2013{\natexlab{d}}, in {Proceedings, 33rd
  International Cosmic Ray Conference (ICRC2013): Rio de Janeiro, Brazil, July
  2-9, 2013}

\bibitem[{Aartsen {et~al.}(2014{\natexlab{a}})}]{Aartsen:2013bfa}
---. 2014{\natexlab{a}},
  \href{http://dx.doi.org/10.1016/j.nima.2013.10.074}{\JournalTitle{Nucl.
  Instrum. Meth.}, A736, 143}

\bibitem[{Aartsen {et~al.}(2014{\natexlab{b}})}]{Aartsen:2014cva}
---. 2014{\natexlab{b}},
  \href{http://dx.doi.org/10.1088/0004-637X/796/2/109}{\JournalTitle{Astrophys.
  J.}, 796, 109}

\bibitem[{Aartsen {et~al.}(2015{\natexlab{a}})}]{Aartsen:2014aqy}
---. 2015{\natexlab{a}},
  \href{http://dx.doi.org/10.1088/2041-8205/805/1/L5}{\JournalTitle{Astrophys.
  J.}, 805, L5}

\bibitem[{Aartsen {et~al.}(2015{\natexlab{b}})}]{Aartsen:2015wto}
---. 2015{\natexlab{b}},
  \href{http://dx.doi.org/10.1088/0004-637X/807/1/46}{\JournalTitle{Astrophys.
  J.}, 807, 46}

\bibitem[{Aartsen {et~al.}(2016{\natexlab{a}})}]{Aartsen:2016ngq}
---. 2016{\natexlab{a}},
  \href{http://dx.doi.org/10.1103/PhysRevLett.117.241101,
  10.1103/PhysRevLett.119.259902}{\JournalTitle{Phys. Rev. Lett.}, 117,
  241101}, [Erratum: Phys. Rev. Lett.119,no.25,259902(2017)]

\bibitem[{Aartsen {et~al.}(2016{\natexlab{b}})}]{Aartsen:2016tpb}
---. 2016{\natexlab{b}},
  \href{http://dx.doi.org/10.3847/2041-8205/824/2/L28}{\JournalTitle{Astrophys.
  J.}, 824, L28}

\bibitem[{Aartsen {et~al.}(2017{\natexlab{a}})}]{Aartsen:2016oji}
---. 2017{\natexlab{a}},
  \href{http://dx.doi.org/10.3847/1538-4357/835/2/151}{\JournalTitle{Astrophys.
  J.}, 835, 151}

\bibitem[{Aartsen {et~al.}(2017{\natexlab{b}})}]{Aartsen:2016nxy}
---. 2017{\natexlab{b}},
  \href{http://dx.doi.org/10.1088/1748-0221/12/03/P03012}{\JournalTitle{JINST},
  12, P03012}

\bibitem[{Aartsen {et~al.}(2018{\natexlab{a}})}]{Aartsen:2017zvw}
---. 2018{\natexlab{a}},
  \href{http://dx.doi.org/10.3847/1538-4357/aab4f8}{\JournalTitle{Astrophys.
  J.}, 857, 117}

\bibitem[{Aartsen {et~al.}(2018{\natexlab{b}})}]{Aartsen:2018vtx}
---. 2018{\natexlab{b}},
  \href{http://dx.doi.org/10.1103/PhysRevD.98.062003}{\JournalTitle{Phys.
  Rev.}, D98, 062003}

\bibitem[{Aartsen {et~al.}(2018{\natexlab{c}})}]{IceCube:2018dnn}
---. 2018{\natexlab{c}},
  \href{http://dx.doi.org/10.1126/science.aat1378}{\JournalTitle{Science}, 361,
  eaat1378}

\bibitem[{Aartsen {et~al.}(2018{\natexlab{d}})}]{IceCube:2018cha}
---. 2018{\natexlab{d}},
  \href{http://dx.doi.org/10.1126/science.aat2890}{\JournalTitle{Science}, 361,
  147}

\bibitem[{Aartsen {et~al.}(2019{\natexlab{a}})}]{Aartsen:2018ywr}
---. 2019{\natexlab{a}},
  \href{http://dx.doi.org/10.1140/epjc/s10052-019-6680-0}{\JournalTitle{Eur.
  Phys. J.}, C79, 234}

\bibitem[{Aartsen {et~al.}(2019{\natexlab{b}})}]{Aartsen:2019fau}
---. 2019{\natexlab{b}}, \href{http://arxiv.org/abs/1910.08488}{{\sffamily
  arXiv:1910.08488 [astro-ph.HE]}}

\bibitem[{Abbasi {et~al.}(2009)}]{Abbasi:2008aa}
Abbasi, R., {et~al.} 2009,
  \href{http://dx.doi.org/10.1016/j.nima.2009.01.001}{\JournalTitle{Nucl.
  Instrum. Meth.}, A601, 294}

\bibitem[{Abbasi {et~al.}(2010)}]{Abbasi:2010vc}
---. 2010,
  \href{http://dx.doi.org/10.1016/j.nima.2010.03.102}{\JournalTitle{Nucl.
  Instrum. Meth.}, A618, 139}

\bibitem[{Abbasi {et~al.}(2011)}]{Abbasi:2010rd}
---. 2011,
  \href{http://dx.doi.org/10.1088/0004-637X/732/1/18}{\JournalTitle{Astrophys.
  J.}, 732, 18}

\bibitem[{Abdullah {et~al.}(2019)Abdullah, Dutta, Ghosh, \&
  Li}]{Abdullah:2019ofw}
Abdullah, M., Dutta, B., Ghosh, S., \& Li, T. 2019,
  \href{http://dx.doi.org/10.1103/PhysRevD.100.115006}{\JournalTitle{Phys.
  Rev.}, D100, 115006}

\bibitem[{Achterberg {et~al.}(2006)}]{Achterberg:2006md}
Achterberg, A., {et~al.} 2006,
  \href{http://dx.doi.org/10.1016/j.astropartphys.2006.06.007}{\JournalTitle{Astropart.
  Phys.}, 26, 155}

\bibitem[{Ahrens {et~al.}(2004)}]{Ahrens:2003fg}
Ahrens, J., {et~al.} 2004,
  \href{http://dx.doi.org/10.1016/j.nima.2004.01.065}{\JournalTitle{Nucl.
  Instrum. Meth.}, A524, 169}

\bibitem[{Anchordoqui \& Antoniadis(2019)}]{Anchordoqui:2018ssd}
Anchordoqui, L.~A., \& Antoniadis, I. 2019,
  \href{http://dx.doi.org/10.1016/j.physletb.2019.02.003}{\JournalTitle{Phys.
  Lett.}, B790, 578}

\bibitem[{Anchordoqui {et~al.}(2018)Anchordoqui, Barger, Learned, Marfatia, \&
  Weiler}]{Anchordoqui:2018ucj}
Anchordoqui, L.~A., Barger, V., Learned, J.~G., Marfatia, D., \& Weiler, T.~J.
  2018, \href{http://dx.doi.org/10.31526/LHEP.1.2018.03}{\JournalTitle{LHEP},
  1, 13}

\bibitem[{Askar'yan(1962)}]{Askaryan:1962hbi}
Askar'yan, G.~A. 1962, \JournalTitle{Sov. Phys. JETP}, 14, 441, [Zh. Eksp.
  Teor. Fiz.41,616(1961)]

\bibitem[{Borah {et~al.}(2019)Borah, Dasgupta, Dey, \& Tomar}]{Borah:2019ciw}
Borah, D., Dasgupta, A., Dey, K., \& Tomar, G. 2019,
  \href{http://arxiv.org/abs/1907.02740}{{\sffamily arXiv:1907.02740 [hep-ph]}}

\bibitem[{Braun {et~al.}(2010)Braun, Baker, Dumm, Finley, Karle, \&
  Montaruli}]{Braun:2009wp}
Braun, J., Baker, M., Dumm, J., {et~al.} 2010,
  \href{http://dx.doi.org/10.1016/j.astropartphys.2010.01.005}{\JournalTitle{Astropart.
  Phys.}, 33, 175}

\bibitem[{{Braun} {et~al.}(2010){Braun}, {Baker}, {Dumm}, {Finley}, {Karle}, \&
  {Montaruli}}]{2010APh....33..175B}
{Braun}, J., {Baker}, M., {Dumm}, J., {et~al.} 2010,
  \href{http://dx.doi.org/10.1016/j.astropartphys.2010.01.005}{\JournalTitle{Astroparticle
  Physics}, 33, 175}

\bibitem[{Carver(2019)}]{Carver:2019jcd}
Carver, T. 2019, in {36th International Cosmic Ray Conference (ICRC 2019)
  Madison, Wisconsin, USA, July 24-August 1, 2019}

\bibitem[{Chauhan \& Mohanty(2019)}]{Chauhan:2018lnq}
Chauhan, B., \& Mohanty, S. 2019,
  \href{http://dx.doi.org/10.1103/PhysRevD.99.095018}{\JournalTitle{Phys.
  Rev.}, D99, 095018}

\bibitem[{Cherry \& Shoemaker(2019)}]{Cherry:2018rxj}
Cherry, J.~F., \& Shoemaker, I.~M. 2019,
  \href{http://dx.doi.org/10.1103/PhysRevD.99.063016}{\JournalTitle{Phys.
  Rev.}, D99, 063016}

\bibitem[{Chipman {et~al.}(2019)Chipman, Diesing, Reno, \&
  Sarcevic}]{Chipman:2019vjm}
Chipman, S., Diesing, R., Reno, M.~H., \& Sarcevic, I. 2019,
  \href{http://dx.doi.org/10.1103/PhysRevD.100.063011}{\JournalTitle{Phys.
  Rev.}, D100, 063011}

\bibitem[{Cline {et~al.}(2019)Cline, Gross, \& Xue}]{Cline:2019snp}
Cline, J.~M., Gross, C., \& Xue, W. 2019,
  \href{http://dx.doi.org/10.1103/PhysRevD.100.015031}{\JournalTitle{Phys.
  Rev.}, D100, 015031}

\bibitem[{Collins {et~al.}(2019)Collins, Bhupal~Dev, \& Sui}]{Collins:2018jpg}
Collins, J.~H., Bhupal~Dev, P.~S., \& Sui, Y. 2019,
  \href{http://dx.doi.org/10.1103/PhysRevD.99.043009}{\JournalTitle{Phys.
  Rev.}, D99, 043009}

\bibitem[{Connolly {et~al.}(2018)Connolly, Allison, \&
  Banerjee}]{Connolly:2018ewv}
Connolly, A., Allison, P., \& Banerjee, O. 2018,
  \href{http://arxiv.org/abs/1807.08892}{{\sffamily arXiv:1807.08892
  [astro-ph.HE]}}

\bibitem[{Cremonesi {et~al.}(2019)}]{Cremonesi:2019zzc}
Cremonesi, L., {et~al.} 2019,
  \href{http://dx.doi.org/10.1088/1748-0221/14/08/P08011}{\JournalTitle{JINST},
  14, P08011}

\bibitem[{de~Vries \& Prohira(2019)}]{deVries:2019gzs}
de~Vries, K.~D., \& Prohira, S. 2019,
  \href{http://dx.doi.org/10.1103/PhysRevLett.123.091102}{\JournalTitle{Phys.
  Rev. Lett.}, 123, 091102}

\bibitem[{Dudas {et~al.}(2018)Dudas, Gherghetta, Kaneta, Mambrini, \&
  Olive}]{Dudas:2018npp}
Dudas, E., Gherghetta, T., Kaneta, K., Mambrini, Y., \& Olive, K.~A. 2018,
  \href{http://dx.doi.org/10.1103/PhysRevD.98.015030}{\JournalTitle{Phys.
  Rev.}, D98, 015030}

\bibitem[{Esmaili \& Farzan(2019)}]{Esmaili:2019pcy}
Esmaili, A., \& Farzan, Y. 2019,
  \href{http://dx.doi.org/10.1088/1475-7516/2019/12/017}{\JournalTitle{JCAP},
  1912, 017}

\bibitem[{Esteban {et~al.}(2019)Esteban, Lopez-Pavon, Martinez-Soler, \&
  Salvado}]{Esteban:2019hcm}
Esteban, I., Lopez-Pavon, J., Martinez-Soler, I., \& Salvado, J. 2019,
  \href{http://arxiv.org/abs/1905.10372}{{\sffamily arXiv:1905.10372 [hep-ph]}}

\bibitem[{Fox {et~al.}(2018)Fox, Sigurdsson, Shandera, Mészáros, Murase,
  Mostafá, \& Coutu}]{Fox:2018syq}
Fox, D.~B., Sigurdsson, S., Shandera, S., {et~al.} 2018,
  \JournalTitle{Submitted to: Phys. Rev. D},
  \href{http://arxiv.org/abs/1809.09615}{{\sffamily arXiv:1809.09615
  [astro-ph.HE]}}

\bibitem[{Gaisser {et~al.}(1995)Gaisser, Halzen, \& Stanev}]{Gaisser:1994yf}
Gaisser, T.~K., Halzen, F., \& Stanev, T. 1995,
  \href{http://dx.doi.org/10.1016/0370-1573(95)00003-Y}{\JournalTitle{Phys.
  Rept.}, 258, 173}, [Erratum: Phys. Rept.271,355(1996)]

\bibitem[{Gorham {et~al.}(2009)}]{Gorham:2008dv}
Gorham, P.~W., {et~al.} 2009,
  \href{http://dx.doi.org/10.1016/j.astropartphys.2009.05.003}{\JournalTitle{Astropart.
  Phys.}, 32, 10}

\bibitem[{Gorham {et~al.}(2016)}]{Gorham:2016zah}
---. 2016,
  \href{http://dx.doi.org/10.1103/PhysRevLett.117.071101}{\JournalTitle{Phys.
  Rev. Lett.}, 117, 071101}

\bibitem[{Gorham {et~al.}(2018{\natexlab{a}})}]{Allison:2018cxu}
---. 2018{\natexlab{a}},
  \href{http://dx.doi.org/10.1103/PhysRevD.98.022001}{\JournalTitle{Phys.
  Rev.}, D98, 022001}

\bibitem[{Gorham {et~al.}(2018{\natexlab{b}})}]{Gorham:2018ydl}
---. 2018{\natexlab{b}},
  \href{http://dx.doi.org/10.1103/PhysRevLett.121.161102}{\JournalTitle{Phys.
  Rev. Lett.}, 121, 161102}

\bibitem[{Greisen(1966)}]{Greisen:1966jv}
Greisen, K. 1966,
  \href{http://dx.doi.org/10.1103/PhysRevLett.16.748}{\JournalTitle{Phys. Rev.
  Lett.}, 16, 748}

\bibitem[{Haack \& Wiebusch(2018)}]{Haack:2017dxi}
Haack, C., \& Wiebusch, C. 2018,
  \href{http://dx.doi.org/10.22323/1.301.1005}{\JournalTitle{PoS}, ICRC2017,
  1005}

\bibitem[{Heurtier {et~al.}(2019{\natexlab{a}})Heurtier, Kim, Park, \&
  Shin}]{Heurtier:2019rkz}
Heurtier, L., Kim, D., Park, J.-C., \& Shin, S. 2019{\natexlab{a}},
  \href{http://dx.doi.org/10.1103/PhysRevD.100.055004}{\JournalTitle{Phys.
  Rev.}, D100, 055004}

\bibitem[{Heurtier {et~al.}(2019{\natexlab{b}})Heurtier, Mambrini, \&
  Pierre}]{Heurtier:2019git}
Heurtier, L., Mambrini, Y., \& Pierre, M. 2019{\natexlab{b}},
  \href{http://dx.doi.org/10.1103/PhysRevD.99.095014}{\JournalTitle{Phys.
  Rev.}, D99, 095014}

\bibitem[{Hooper {et~al.}(2019)Hooper, Wegsman, Deaconu, \&
  Vieregg}]{Hooper:2019ytr}
Hooper, D., Wegsman, S., Deaconu, C., \& Vieregg, A. 2019,
  \href{http://dx.doi.org/10.1103/PhysRevD.100.043019}{\JournalTitle{Phys.
  Rev.}, D100, 043019}

\bibitem[{Hoover {et~al.}(2010)}]{Hoover:2010qt}
Hoover, S., {et~al.} 2010,
  \href{http://dx.doi.org/10.1103/PhysRevLett.105.151101}{\JournalTitle{Phys.
  Rev. Lett.}, 105, 151101}

\bibitem[{Huang(2018)}]{Huang:2018als}
Huang, G.-y. 2018,
  \href{http://dx.doi.org/10.1103/PhysRevD.98.043019}{\JournalTitle{Phys.
  Rev.}, D98, 043019}

\bibitem[{Motloch {et~al.}(2017)Motloch, Alvarez-Muñiz, Privitera, \&
  Zas}]{Motloch:2016yic}
Motloch, P., Alvarez-Muñiz, J., Privitera, P., \& Zas, E. 2017,
  \href{http://dx.doi.org/10.1103/PhysRevD.95.043004}{\JournalTitle{Phys.
  Rev.}, D95, 043004}

\bibitem[{Prohira {et~al.}(2018)}]{Prohira:2018mmv}
Prohira, S., {et~al.} 2018,
  \href{http://dx.doi.org/10.1103/PhysRevD.98.042004}{\JournalTitle{Phys.
  Rev.}, D98, 042004}

\bibitem[{Romero-Wolf {et~al.}(2019)}]{Romero-Wolf:2018zxt}
Romero-Wolf, A., {et~al.} 2019,
  \href{http://dx.doi.org/10.1103/PhysRevD.99.063011}{\JournalTitle{Phys.
  Rev.}, D99, 063011}

\bibitem[{Safa {et~al.}(2020)Safa, Pizzuto, Argüelles, Halzen, Hussain,
  Kheirandish, \& Vandenbroucke}]{Safa:2019ege}
Safa, I., Pizzuto, A., Argüelles, C.~A., {et~al.} 2020,
  \href{http://dx.doi.org/10.1088/1475-7516/2020/01/012}{\JournalTitle{JCAP},
  2001, 012}

\bibitem[{Safa {et~al.}(2019)}]{Safa:2019icrc}
Safa, I., {et~al.} 2019, in {36th International Cosmic Ray Conference (ICRC
  2019) Madison, Wisconsin, USA, July 24-August 1, 2019}

\bibitem[{Schumacher(2019)}]{Schumacher:2019qdx}
Schumacher, L. 2019,
  \href{http://dx.doi.org/10.1051/epjconf/201920702010}{\JournalTitle{EPJ Web
  Conf.}, 207, 02010}

\bibitem[{Shoemaker {et~al.}(2019)Shoemaker, Kusenko, Munneke, Romero-Wolf,
  Schroeder, \& Siegert}]{Shoemaker:2019xlt}
Shoemaker, I.~M., Kusenko, A., Munneke, P.~K., {et~al.} 2019,
  \href{http://arxiv.org/abs/1905.02846}{{\sffamily arXiv:1905.02846
  [astro-ph.HE]}}

\bibitem[{Venters {et~al.}(2019)Venters, Reno, Krizmanic, Anchordoqui, Guépin,
  \& Olinto}]{Venters:2019xwi}
Venters, T.~M., Reno, M.~H., Krizmanic, J.~F., {et~al.} 2019,
  \href{http://arxiv.org/abs/1906.07209}{{\sffamily arXiv:1906.07209
  [astro-ph.HE]}}

\bibitem[{Zas(2018)}]{Zas:2017xdj}
Zas, E. 2018, \href{http://dx.doi.org/10.22323/1.301.0972}{\JournalTitle{PoS},
  ICRC2017, 972}, [,64(2017)]

\bibitem[{Zatsepin \& Kuzmin(1966)}]{Zatsepin:1966jv}
Zatsepin, G.~T., \& Kuzmin, V.~A. 1966, \JournalTitle{JETP Lett.}, 4, 78,
  [Pisma Zh. Eksp. Teor. Fiz.4,114(1966)]

\end{thebibliography}

\end{document}